\documentclass[twocolumn,showpacs,preprintnumbers,amsmath,amssymb]{revtex4-1}

\usepackage{placeins}
\usepackage{graphicx}

\usepackage{amssymb}
\usepackage{amsthm}
\newtheorem{defn}{Definition}

\usepackage{amsmath,amssymb,amsfonts,latexsym,cancel}
\usepackage{rawfonts}
\usepackage{pictexwd}
\usepackage[all]{xy}
\newtheorem {theorem}{Theorem}
\newtheorem {lemma}{Lemma}

\begin{document}


\title{Virus and Warning Spread in Dynamical Networks}

\author{Carlos Rodr{\'i}guez-Lucatero}
\email{crodriguez@correo.cua.uam.mx \& rbernal@correo.cua.uam.mx }
\affiliation{Departamento de Tecnolog\'{\i}as de la Informaci\'on, Universidad Aut\'onoma Metropolitana-Cuajimalpa,
Av. Constituyentes 1056,
Col.Lomas Altas, M\'exico, D. F.,
C.P. 11950, M\'exico}
\author{Roberto Bernal-Jaquez }
\affiliation{ Departamento de Matem\'aticas Aplicadas y Sistemas,\\
 Universidad Aut\'onoma Metropolitana-Cuajimalpa,
 Artificios 40, Col. Hidalgo, Delegaci\'on \'Alvaro Obreg\'on,
 M\'exico, D.F  01120, M\'exico }

\begin{abstract}
Recent work on information survival in sensor and human P2P networks, try to study the datum preservation or the virus spreading in a network under the dynamical system approach. Some interesting solutions propose to use non-linear dynamical systems and fixed point stability theorems, providing closed form formulas that depend on the largest eigenvalue of the dynamic system matrix. Given that in a the Web there can be messages from one place to another, and that this messages can be, with some probability, new unclassified virus warning messages as well as worms or other kind of virus, the sites can be infected very fast. The question to answer is how and when a network infection can become global and how it can be controlled or at least how to stabilize his spreading in such a way that it becomes confined below a fixed portion of the network.
In this paper, we try to make a step ahead in this direction and apply classic results of the dynamical systems theory to model the behaviour of a network where warning messages and virus spread. 

\end{abstract}


\maketitle


\section{Introduction}
Recently with the constant augmentation in the number of internet users as  well as the growth in the complexity of such networks,  new security problems have appeared in the scene and there is a lack of adequate security methods for facing attacks under this new setting. These new environments are for instance  the P2P networks, sensor networks, social nets or wireless networks, where information is to be stored, generated and retrieved. So under this new environments it can be very important to study and model how the information is spread or how to keep the spreading of a virus under control in such a way that the information still being useful under these vulnerable circumstances. In \cite{Deepa1}  it is studied the problem of information survival threshold in sensor and P2P networks, modeling the problem as a non-linear dynamical system and using fixed point stability theorems, and obtain a closed form solution that depends on an additional parameter, the largest eigenvalue of the dynamical system matrix. In the sensor networks for instance, the nodes can loss their communications links and the nodes can stop working because of system failure produced by a virus infection and  quarantine process or a system maintenance procedure. Under such conditions they try to answer the following question:

{\bf PROBLEM:} \emph{Under what conditions a datum can survive in a sensor network?} 
Given that the nodes as well as the links can fail with some probability the obvious model can be a Markov chain, but such a model can grow in complexity very quickly because the number of possible states becomes $3^{N}$ where $N$ is the number of states. To avoid this mathematical problem, one alternative is to model the system as a non-linear dynamical system.
Recently they have appeared in the conferences and journal articles some very interesting and relevant research articles about the virus spread behaviour in a P2P network or in scale free nets such as the Web. In \cite{Kempe1} the authors study the communication mechanisms for gossip based protocols. Another very recent and interesting work on how to distribute antidotes for controlling  the epidemics spread is presented in \cite{chayes1}. In this research the authors analyse the problem under the approach of contact processes \cite{Durrett1} on a finite graph and obtain very interesting and rigorous results. Concerning the properties that arise in the random graphs, such as the existence of a \emph{giant component}, \emph{percolation} phenomena, node degree distribution and \emph{small world} phenomena, and that are the base of many recent works on virus spread on networks, we can mention \cite{Barab0,Barab1,Barab2} as well as \cite{Alon1} \cite{Bollobas1} \cite{Falou1} and \cite{Radicchi1}. Concerning the subject on mathematical modeling of epidemic spreading we should mention the outstanding work done by Romualdo Pastor-Satorras and Alessandro Vespignani in \cite{Satorras1,Satorras2,Satorras3}.    
In the present work we will take as source of inspiration \cite{Deepa1}. 
In \cite{Deepa1} the authors implement some experiments on several real sensor and P2P networks (from Intel, MIT, Gnutella, and others) to show the accuracy of their method.
In this work it is claimed that their method is not only applicable to sensor nets but it is also applicable to many more settings where a piece of information may be replicated across faulty links and faulty
nodes. The authors establish a survivability condition that produce a bound in the design of distributed
systems, allowing to: 

\begin{itemize}

\item Estimate the most economic retransmission
rates for a datum to survive in a sensor network.

\item Decide which nodes can be removed while still remaining above the survivability
threshold in a sensor network.

\item Drive a \emph{virus} as \emph{datum} to extinction for anti-virus protection, 
by deciding how often to quarantine nodes and how long they should be
kept down.

\item Propagate and maintain information (news, rumors, etc.)

\end{itemize}

So, this work is closely related with the areas of gossip based
protocols, epidemiology and computer security. Gossip-based
protocols on networks, whose related graphs have dynamic presence of nodes, that keep some level of state consistency have been proposed in \cite{Kempe1}. The basic underlying idea of the gossip protocols is that at each time step each node $i$ chose to communicate with a node $j$ generally following a random rule, exchanging information during a period of time, spreading it in the system in the same way 
as the virus are spread. A fundamental issue in this kind of protocols is how the underlying gossip low level mechanism affects the ability to design efficient high level gossip protocol algorithms.
In \cite{Kempe1} the authors show a fundamental limitation on the power of the commonly used uniform
gossip mechanism for solving nearest-resource location problems. They show as well that very efficient protocols for this
problem can be designed using a non-uniform spatial gossip mechanism. The gossip-based distributed protocol algorithms obtained in \cite{Kempe1} for complex problems for a set of nodes in Euclidean space are implemented by constructing an approximated minimum spanning tree. 

\subsubsection{Previous proposed mathematical model of virus spread.} \label{Virus2}

With the increasing importance and presence of sensor as well as P2P networks, networks have a high level of congestion and because of that the theory concerning information survivability becomes very important. 
One source of inspiration for mathematical modeling problems of information survivability in this kind of networks is the \emph{epidemiology}.  


The \emph{epidemiology} community has developed several stochastic models for studying the
spread and die-out of diseases in a population. The two most relevant ones are the SIR and SIS models. Both are stochastic models of the spread of disease through a population, where the \emph{susceptible}
nodes can get \emph{infected} on contact with \emph{infected}
neighbors. The \emph{infected} hosts eventually die (in the SIR model) or recover and become \emph{susceptible} again (in the SIS model). The point of view adopted in \cite{Deepa1}
was the SIS model. Under this model  a node is \emph{susceptible} to a data item when it is online and
under normal operation. When the nodes start to fail, they become \emph{immune} during their failure, and later they become \emph{susceptible} again when they are back online. Some results obtained in \cite{Deepa1} are very useful to analyze the survival of a infection in a population, based on the graph theory results similar to the ones mentioned in \cite{Barab1,Barab2,Alon1}. In computer security one of the important issues that have been studied under SIS and SIR infection spreading mathematical models are the virus propagation as well as worms on Internet, from where, the exponential spread of them and the \emph{epidemic thresholds} can be estimated \cite{Satorras1},\cite{Satorras2}, \cite{Satorras3}.
Let us suppose that we have a sensor/P2P/social network of $N$ nodes (sensors or computers or people) and $E$
directed links between them. Let us also assume that we take very small discrete timesteps of size $\Delta t$ where $\Delta t \rightarrow 0$. The survivability results in \cite{Deepa1} apply equally well to continuous systems. Within a $\Delta t$ time interval, each node $i$ has probabiity $r_{i}$ of trying to broadcast its information every time step, and each link $i \rightarrow j$ has a probability $\beta_{i,j}$ of being \emph{up}, and thus correctly propagating the information to node $j$. Each node $i$ also has a node failure probability $\delta_{i} > 0$ (e.g., due to battery failure
in sensors). Every dead node $j$ has a rate $\gamma_{j}$ of returning to the \emph{up} state, but without any information
in its memory (e.g., due to the periodic replacement of dead batteries). These and other symbols are listed
in Table 1. \\
\\
{\scriptsize
$
\begin{array}{|l|l|}
      \hline
			   \mbox{Symbol} & \mbox{Description}  \\
			\hline
			  N            & \mbox{Number of nodes in a network}   \\
			  \beta_{ij}  & \mbox{Probability that the link} \\
			  ~~~~~         & i \rightarrow j \mbox{is up}  \\
			  \delta_{i} & \mbox{Death rate: Probability that node~} i \mbox{~dies}  \\
			  \gamma_{i} & \mbox{Resurrection rate:}\\
			  ~~~~         & \mbox{Probability that node~} i \mbox{~comes back up} \\
			  r_{i}      & \mbox{Retransmission rate:}\\ 
			  ~~~~~        & \mbox{Probability that node} i \mbox{~broadcasts} \\
			 \hline
			 p_{i}(t)    &\mbox{Probability that node}\\
			 ~~~~~         & i \mbox{~ is alive at time} t \mbox{~and has info} \\
			 q_{i}(t)    &\mbox{Probability that node}\\
			 ~~~~          & i \mbox{ ~ is alive at time} t \mbox{~but without info} \\
			 1-p_{i}(t)-q_{i}(t) & \mbox{Probability that node} i \mbox{~is dead}\\
			 \zeta_{i}   & \mbox{Probability that node} i \mbox{~does}\\
			 ~~~~          & \mbox{not receive info from}\\ 
			 ~~~~          & \mbox{any of its neighbors at time~} t \\
			 \vec{p}(t), \vec{q}(t)   & \mbox{Probability column vectors} \\
			 f:\Re^{2N} \rightarrow \Re^{2N} &\mbox{Function representing a dynamical system}\\
			 \nabla (f)    & \mbox{The Jacobian matrix of~} f(.) \\
			 S             & \mbox{The~} N \times N \mbox{~system matrix} \\
			 \lambda_{S}   & \mbox{An eigenvalue of the~} S \mbox{matrix}\\
			 \lambda_{1,S}   & \mbox{The largest in magnitude}\\
			 ~~~~            & \mbox{eigenvalue of the} S \mbox{matrix}\\
			 s=|\lambda_{1,S}| & \mbox{Survivability score = Magnitude of} \lambda_{1,S}\\
			 
			 \hline
\end{array}
$
}\\
\\
This system can be modeled as a Markov chain,
where each node can be in one of three states: \emph{Has Info}, \emph{No Info} or \emph{Dead}, with transitions between them as shown in Diagram 1. The full state of the system
at any instant consists of $N$ such states, one for each node. Thus, there are $3^{N}$ system states. Transitions out of the current system state depend only on the current state and not on any previous states; thus it is a Markov chain.\\ 
The next graph represent the transition that take place in each node.
\\
\unitlength=0.5mm
\special{em:linewidth 0.4pt}
\linethickness{0.4pt}
\begin{picture}(200.00,200.00)
\put(50.00,100.00){\circle{34.00}} 
\put(120.00,100.00){\circle{34.00}} 
\put(100.00,70.00){\circle{34.00}} 
\bezier{360}(35.00,100.00)(45.00,170.00)(50.00,115.00)
\put(37,110){\vector(-1,-4){2}} 
\bezier{364}(55.00,115.00)(84.00,169.00)(118.00,117.00)
\put(56,120){\vector(-1,-3){2}} 
\bezier{365}(55.00,86.00)(60.00,50.00)(86.00,70.00)
\put(85,70){\vector(1,1){2}} 
\bezier{156}(130.00,110.00)(150.00,120.00)(130.00,90.00)
\put(132,112){\vector(-1,-1){2}} 
\bezier{157}(90.00,80.00)(96.00,100.00)(106.00,95.00)
\put(90,80){\vector(-1,-1){2}} 
\bezier{132}(100.00,56.00)(130.00,40.00)(112.00,65.00)
\put(114,63){\vector(-1,1){2}} 
\bezier{137}(115.00,70.00)(125.00,60.00)(128.00,90.00)
\put(127,83){\vector(1,4){2}} 
\put(145.00,80.00){\makebox(0,0)[cc]{Resurrected}} 
\put(145.00,70.00){\makebox(0,0)[cc]{$\gamma_{i}$}} 
\put(143.00,60.00){\makebox(0,0)[cc]{Prob $1-p_{i}-q_{i}$}} 
\put(100.00,48.00){\makebox(0,0)[cc]{$1-\gamma_{i}$}} 
\put(47.00,70.00){\makebox(0,0)[cc]{Dies}} 
\put(47.00,60.00){\makebox(0,0)[cc]{$\delta_{i}$}} 
\put(50.00,99.00){\makebox(0,0)[cc]{Has}}
\put(50.00,93.00){\makebox(0,0)[cc]{Info}}
\put(25.00,92.00){\makebox(0,0)[cc]{Prob $p_{i}$}}
\put(30.00,140.00){\makebox(0,0)[cc]{$1-\delta_{i}$}} 
\put(86.00,102.00){\makebox(0,0)[cc]{Dies}} 
\put(86.00,92.00){\makebox(0,0)[cc]{$\delta_{i}$}} 
\put(83.00,145.00){\makebox(0,0)[cc]{Receives Info}}
\put(83.00,135.00){\makebox(0,0)[cc]{$1-\zeta_{i}(t)$}}
\put(120.00,99.00){\makebox(0,0)[cc]{No}}
\put(120.00,93.00){\makebox(0,0)[cc]{Info}}
\put(135.00,122.00){\makebox(0,0)[cc]{$\zeta_{i}(t)-\delta_{i}$}}
\put(145.00,92.00){\makebox(0,0)[cc]{Prob $q_{i}$}}
\put(100.00,70.00){\makebox(0,0)[cc]{Dead}}
\put(100,20){\makebox(0,0)[cc]{Diagram 1: Transitions on each node}}
\end{picture}

It can be pointed out that there is an absorbing set of states where no node is in \emph{Has Info} state.
Under such circumstances the information dies with probability $1$ as $t \rightarrow \infty$.
Some combination of parameters lend the system quickly to this state and some other combination does not so in practice the datum \emph{survives} for some parameter combination.
The question is: under what conditions does the
information survive for a long time, and when will
the information die out quickly?
Let $\overline{C}(t)$ denote the
expected number of \emph{carriers} (nodes in \emph{Has Info}
state) at time $t$. In general, $\overline{C}(t)$  decays exponentially,
polynomially or logarithmically (with expected
extinction time comparable to or larger than
the age of the universe for large graphs), depending
on the system being below, at or above a threshold \cite{Durrett1}. Let us focus on 
the \emph{fast extinction} case where $\overline{C} (t)$ decays exponentially.

{\defn
\emph{Fast extinction} is the setting where the number of carriers $\overline{C}(t)$ 
decays exponentially over time ($\overline{C} (t) \propto c^{-t}, c > 1$)

}
Now, the problem can be formally stated as follows:
{\bf PROBLEM:}
\begin{itemize}

\item Given: the network topology (link \emph{up} probabilities) $\beta_{ij}$ the retransmission
rates $r_{i}$, the resurrection rates $\gamma_{i}$ and the death rates ($\delta_{i}$ $i=1 \ldots N, j=1 \ldots N$)

\item Find the condition under which a datum
will suffer \emph{fast extinction}.

\end{itemize}


To simplify the problem and to avoid dependencies
on starting conditions, we consider the case where
all nodes are initially in the \emph{have info} state.

\subsubsection{Main Idea}

Solving this problem for the full Markov chain requires $3^{N}$ variables and is thus intractable, even for
moderate-sized networks. Exact values for the \emph{fast extinction} threshold are unavailable even for simpler versions of this problem. The main contribution of \cite{Deepa1}
is an accurate approximation, using a non-linear
dynamical system of only $N$ variables. The heart of
their approximation is to consider the states of
the two different nodes to be mutually independent.
Let the probability of node $i$ being in the \emph{Has Info} and \emph{No Info} states at time $t$ be $p_{i} (t)$ and $q_{i} (t)$ respectively. Thus, the probability of its being dead is $(1-p_{i} (t)-q_{i} (t))$. Starting from state
\emph{No Info} at time $t-1$, node $i$ can acquire this information
(and move to state \emph{Has Info}) if it receives a
communication from some other node $j$. Let $\zeta_{i}(t)$ be
the probability that node $i$ does not receive the information
from any of its neighbors. Then, assuming
the neighbor's states are independent:\\

\begin{equation} \label{indep}
	\zeta_{i}(t)=\prod_{j=1}^{N} (1-r_{j}\beta_{ji}p_{j}(t-1))
\end{equation}

For each node $i$, we can use the transition matrix in
Diagram 1 to write down the probabilities of being in
each state at time $t$, given the probabilities at time $t-1$ (recall that we use very small time steps $\Delta t$, and so we can neglect second-order terms). Thus:

\begin{equation}\label{probp}
	p_{i}(t)=p_{i}(t-1)(1-\delta_{i})+q_{i}(t-1)(1-\zeta_{i}(t))
\end{equation}

\begin{equation}\label{probq}
	q_{i}(t)=q_{i}(t-1)(\zeta_{i}(t)-\delta_{i})+(1-p_{i}(t-1)-q_{i}(t-1))\gamma_{i}
\end{equation}

\subsubsection{Main previous Results} \label{virus2}

In  \cite{Deepa1}, experimental results have been obtained under fast extinction conditions that accurately correspond to what their model predicts. The authors claim that the accuracy of their model predictions are due to the \emph{mixing properties} of real networks. For the sake of completeness we will state the main results in \cite{Deepa1} and to show how some of them are obtained because our own results will be obtained by applying the same procedures.

\begin{defn} \label{def1}
Define $S$ to be the $N \times N$ system matrix:\\
\begin{equation*}
S_{ij}= \left \{
\begin{array}{ll}
	1-\delta_{i}  &~~~~\text{if~} i=j \\
	r_{j} \beta_{ji} \frac{\gamma_{i}}{\gamma_{i}+\delta_{i}} & ~~~~\mbox{otherwise}
\end{array}
\right .
\end{equation*}
Let $|\lambda_{1,S}|$ be the magnitude of the largest eigenvalue and $\widehat{C}(t)=\sum_{i=1}^{N} p_{i}(t)$ the expected number of carriers at $t$ of the dynamical system.
\end{defn}

\begin{theorem}
{\bf (Condition for fast extinction)}. Define $s=|\lambda_{1,S}|$ to be the {\bf survivability score} for the system. If $s=|\lambda_{1,S}|<1$, then we have fast extinction in the dynamical system, that is, $\widehat{C}(t)$ decays exponentially quickly over time.
\end{theorem}  
Where $|\lambda_{i,S}|$ is the magnitude of the largest eigenvalue of $S$, being $S$ an $N \times N$ system matrix defined as
$S_{ij}=1-\delta_{i}$ if $i=j$ and $S_{ij}=r_{j}\beta_{ji} \frac{\gamma_{i}}{\gamma_{i}+\delta_{i}}$ otherwise, and being $\widehat{C}(t)=\sum_{i=1}^{N}p_{i}(t)$ the expected number of carriers at time $t$ of the dynamical system.
Two additional results that appears in \cite{Deepa1} are the following

\begin{lemma}\label{lemaDeepa1}
{\bf {Fixed point}}. The values $(p_{i}(t)=0,q_{i}(t)=\frac{\gamma_{i}}{\gamma_{i}+\delta_{i}})$ for all nodes $i$, are a fixed point of the equations (\ref{probp}) and (\ref{probq}). Proved by a simple application of the Equations.
\end{lemma}
  
\begin{theorem}
{\bf (Stability of the fixed point).} The fixed point point of Lemma 1 is asymptotically  if the system is bellow threshold, that is, $s=|\lambda_{1,S}|<1$ 
\end{theorem}  

\begin{lemma}\label{lemaDeepa2}
(From reference [8] of \cite{Deepa1}) Define $\nabla(f)$ (also called the Jacobian matrix) to be a $2N \times 2N$ matrix such that
\begin{equation}
[\nabla(f)]_{ij}=\frac{\partial f_{i}(\vec{v}(t-1))}{\partial \vec{v}_{j}(t-1)}
\end{equation}
where $\vec{v}$ is the concatenation of $\vec{p}$ and $\vec{q}$. Then, if the largest eigenvalue (in magnitude) of $\nabla(f)$ at $\vec{v}_{f}$ (vector $\vec{v}$ valued at the fixed point) is less than $1$ in magnitude, the system is asymptotically stable at $\vec{v}_{f}$. Also, if $f$ is linear and the condition holds, then the dynamical system will exponentially tend to the fixed point irrespective of initial state.
\end{lemma}

In \cite{Deepa1} the authors apply (\ref{lemaDeepa2}) and obtain the following block matrix
\begin{equation}
\nabla(f)|_{\vec{v}_{f}}= \left [
\begin{array}{lll}
S      & |  & 0 \\
\hline 
S_{1}  & | & S_{2}
\end{array}
\right ]
\end{equation} 
The dimensions of each block matrix are $N \times N$ whose elements are

\begin{equation}
S_{ij}= \left \{
\begin{array}{ll}
	1-\delta_{i}  &~~~~\text{if~} i=j \\
	r_{j} \beta_{ji} \frac{\gamma_{i}}{\gamma_{i}+\delta_{i}} & ~~~~\mbox{otherwise.}
\end{array}
\right .
\end{equation}

The others are

\begin{equation}
S_{1ij}= \left \{
\begin{array}{ll}
	1-\delta_{i}  &~~~~\text{if~} i=j \\
	-r_{j} \beta_{ji} \frac{\gamma_{i}}{\gamma_{i}+\delta_{i}} & ~~~~\mbox{otherwise}
\end{array}
\right .
\end{equation}
 
and

\begin{equation}
S_{2ij}= \left \{
\begin{array}{ll}
	1-\gamma_{i}-\delta_{i}  &~~~~\text{if~} i=j \\
	0 & ~~~~\mbox{otherwise}
\end{array}
\right .
\end{equation}

So the question is \emph{how can be obtained the fixed point of the system?}.
In the following paragraph we will sketch, in an alternative way of the used in \cite{Deepa1}, how it can be done.
In dynamical systems theory the \emph{fixed point} is called \emph{equilibrium point} of the system.
In this very point the state probabilities become stable, then $p_{i}(t)=p_{i}(t-1)$ and $q_{i}(t)=q_{i}(t-1)$. Then simplifying the notation by dropping the subindex and the time parameter we can state the following equations system:
\begin{equation}
\begin{array}{l}
p=p \cdot (1-\delta)+ q \cdot (1-\zeta)\\
q=q \cdot (\zeta-\delta) +(1-p-q) \cdot \gamma
\end{array}
\end{equation}

after algebraic simplification it can be obtained the following equations system

\begin{equation}
\begin{array}{l}
-\delta \cdot p  +  (1-\zeta) \cdot q = 0\\
\gamma \cdot p +(\zeta-1-\delta-\gamma) \cdot q  = q  
\end{array}
\end{equation}

Expressing the equations system in matrix form we get
\begin{equation}
\left [
\begin{array}{ll}
-\delta & 1-\zeta \\
-\gamma & \zeta-1-\delta-\gamma
\end{array}
\right ]
\left [
\begin{array}{l}
p \\
q
\end{array}
\right ]
=
\left [
\begin{array}{l}
0 \\
-\gamma
\end{array}
\right ]
\end{equation}

Solving by Cramer's method we obtain

\begin{equation}
\begin{array}{l} \label{fixpoint5}
p =  \frac{\gamma \cdot (1-\zeta)}{\gamma \cdot (1-\zeta)-\delta \cdot(\zeta-1-\delta-\gamma)} \\
q =  \frac{\delta \cdot \gamma }{\gamma \cdot (1-\zeta)-\delta \cdot(\zeta-1-\delta-\gamma)}
\end{array}
\end{equation}

The expressions (\ref{fixpoint5}) can be simplified by observing that the stable state \emph{No Info} is related with the desired \emph{fast extinction} condition that is also related with Markov chain probability condition $(1-\zeta)\rightarrow 0$, that implies $p \rightarrow 0$. Taking into account this fact, we can rewrite (\ref{fixpoint5}) as follows:\\
 
\begin{equation}
\begin{array}{ll} \label{fixpoint6}
p = & \frac{\gamma \cdot (1-\zeta)}{\gamma \cdot (1-\zeta)+\delta \cdot(1-\zeta+\delta+\gamma)}\\
~~  = & 0 \\
q = & \frac{\delta \cdot \gamma }{\gamma \cdot (1-\zeta)-\delta \cdot(\zeta-1-\delta-\gamma)}\\
~~ = & \frac{\delta \cdot \gamma }{\gamma \cdot (1-\zeta)+\delta \cdot(1-\zeta+\delta+\gamma)}\\
~~ = & \frac{\delta \cdot \gamma}{\delta \cdot (\delta+\gamma)}\\
~~ = & \frac{\gamma}{\delta+\gamma}
\end{array}
\end{equation}

\subsection{Markovian Analysis}
In \cite{Deepa1} the Markovian analysis was avoided because of the size of the resulting configuration space ($3^{N}$ where $N$ correspond to the number of nodes and $3$ to number of states in the Markov chain). However, it is interesting to analyse the ergodicity behaviour of the Markov chains associated inside each node. We have done this by the calculation of the corresponding Z-transform of the associated matrix and performing the following steps:

\begin{itemize}

\item We obtain the transition matrix $P$ 
{\footnotesize
\begin{equation}
P=\left(
\begin{array}{ccc}
 (1-\delta ) & 0 & \delta  \\
  (\delta -\zeta ) &  (1-\zeta ) &  \delta  \\
 (1-\gamma ) & 0 &  \gamma 
\end{array}
\right)
\end{equation}
}

\item{ We calculate the Z-transform of the matrix $M=I-zP$, that is
\begin{equation}
M=\left(
\begin{array}{ccc}
 1-z (1-\delta ) & 0 & -z \delta  \\
 z (\delta -\zeta ) & 1-z (1-\zeta ) & -z \delta  \\
 z (-1+\gamma ) & 0 & 1-z \gamma 
\end{array}
\right)
\end{equation}
}
\item {The inverse matrix $M^{-1}$ of $M$ is obtained}

\item {The inverse Z-transform is applied to $M^{-1}$, obtaining 

$H(n)={\cal{Z}}^{-1}\{{M^{-1}}\}$

This equation can be written as a sum of  two matrices \cite{Howard}, $S$ that corresponds to the steady behavior  and $T$ that represents the transient behavior of the Markov Chain, that is  

$H(n)=k_1S+k_2(C_1)^{n}T$,

where $k_1$ and $k_2$ are constants. Based on this procedure we obtained the ergodicity condition, that is, $C_1 <1$ what assures the convergence to a steady state no matter what the initial state was.

All the calculations were made using MATHEMATICA and the ergodicity condition expression obtained is given by

\begin{equation}
\noindent {C_1=\frac{2-\gamma -\delta +\sqrt{\gamma ^2-6 \gamma  \delta +\delta ^2} }{1-\gamma -\delta +2 \gamma  \delta }}.
\end{equation}
}

It's worthy to notice that the ergodicity of the Markov chain depends on the values of the transition probabilities involved. We should also mention that the calculations of this analysis are an additional source of difficulty. The more states the Markov chain have, the more complex become the algebraic manipulation and Z-transforms. So, it is not a surprise that the Markovian analysis was avoided in \cite{Deepa1} and the choice was the dynamical systems approach.  

\end{itemize}

\section{Our proposal}

The model described in subsection \ref{Virus2} and that constitutes the core of the results obtained in \cite{Deepa1} has as main  purpose to estimate the threshold condition under which the propagation of a virus in a P2P network decays exponentially. This last question implies that for keeping the network below the threshold just mentioned, the protocols have to disconnect temporarily some nodes, fix the problem and reboot them. This method is very efficient for stopping the propagation of the virus. In this way the number of virus carriers decays exponentially. Let us assume that at the same time we need to propagate an alarm signal warning about the presence of a worm virus or an antidote \cite{chayes1} in a P2P network. Then in these cases we need that the network operates over the estimated threshold. So we are in a situation where the threshold conditions are  antagonist  and that can happen in a real world setting. Under such circumstances the question is \emph{How to keep a datum and at the same time avoid  the virus spreading?}. 
Our hypothesis is that it will depend on the proportion of virus messages versus warning messages. For this reason we will propose a new model that will take into account this situation.
The previous and the new additional symbols are listed in Table 2.
\\
$
{\scriptsize
\begin{array}{|l|l|}
      \hline
			   \mbox{Symbol} & \mbox{Description}  \\
			\hline
			  N            & \mbox{Number of nodes in a network}   \\
			  \beta_{ij}  & \mbox{Probability that the link} \\
			  ~~~~~         & i \rightarrow j \mbox{is up}  \\
			  \delta_{i} & \mbox{Death rate: Probability that node~} i \mbox{~dies}  \\
			  \gamma_{i} & \mbox{Resurrection rate:}\\
			  ~~~~         & \mbox{Probability that node~} i \mbox{~comes back up} \\
			  r_{i}      & \mbox{Retransmission rate:}\\ 
			  ~~~~~        & \mbox{Probability that node} i \mbox{~broadcasts} \\
			 \hline
			 p_{i}(t)    &\mbox{Probability that node}\\
			 ~~~~~         & i \mbox{~ is infected at time} t \mbox{~and has virus info} \\
			 q_{i}(t)    &\mbox{Probability that node has no Info}\\
			 ~~~~          & i \mbox{ ~ is healthy at time} t \mbox{~but susceptible} \\
			 1-p_{i}(t)-q_{i}(t)-w_{i}(t) & \mbox{Probability that node} i \mbox{~is dead}\\
			 w_{i}(t)    &\mbox{Probability that node has warning Info}\\
			 ~~~~          & i \mbox{ ~ is warned at time} t  \\
			 
			 \zeta_{i}   & \mbox{Probability that node} i \mbox{~does}\\
			 ~~~~          & \mbox{not receive info from}\\ 
			 ~~~~          & \mbox{any of its neighbors at time~} t \\
			 \nu_{i}   & \mbox{Probability that node} i \\
			 ~~~~          & \mbox{receive virus}\\ 
			 1-\nu_{i}   & \mbox{Probability that node} i \\
			 ~~~~          & \mbox{receive a warning}\\ 

			 \chi_{i}   & \mbox{Probability that node} i \\
			 ~~~~          & \mbox{applies vaccin}\\ 
		 
			 \vec{p}(t), \vec{q}(t)   & \mbox{Probability column vectors} \\
			 f:\Re^{2N} \rightarrow \Re^{2N} &\mbox{Function representing a dynamical system}\\
			 \nabla (f)    & \mbox{The Jacobian matrix of~} f(.) \\
			 S             & \mbox{The~} N \times N \mbox{~system matrix} \\
			 \lambda_{S}   & \mbox{An eigenvalue of the~} S \mbox{matrix}\\
			 \lambda_{1,S}   & \mbox{The largest in magnitude}\\
			 ~~~~            & \mbox{eigenvalue of the} S \mbox{matrix}\\
			 s=|\lambda_{1,S}| & \mbox{Survivability score = Magnitude of} \lambda_{1,S}\\
			 
			 \hline
\end{array}
}$
\\
\\This system can be  modeled as well as a Markov chain,
where each node can be in one of three states: \emph{Infected},\emph{Warn Info}, \emph{No Info} or \emph{Dead}, with transitions between them as shown in Diagram 2. The full state of the system
at any instant consists of $N$ such states, one for each node. Therefore, there are $4^{N}$ system states. Transitions out of the current system state depend only on
the current state and not on any previous states; then
it is a Markov chain without memory.
The next graph represent the transitions that take place in each node for our model.\\
\unitlength=0.5mm
\special{em:linewidth 0.4pt}
\linethickness{0.4pt}
\begin{picture}(200.00,200.00)
\put(50.00,100.00){\circle{34.00}} 
\put(120.00,100.00){\circle{34.00}} 
\put(100.00,70.00){\circle{34.00}} 
\put(85.00,122.00){\circle{20.00}} 
\bezier{360}(35.00,100.00)(45.00,170.00)(50.00,115.00)
\put(37,110){\vector(-1,-4){2}} 
\bezier{364}(55.00,115.00)(84.00,169.00)(118.00,117.00)
\put(56,120){\vector(-1,-3){2}} 
\bezier{365}(55.00,86.00)(60.00,50.00)(86.00,70.00)
\put(85,70){\vector(1,1){2}} 
\bezier{156}(130.00,110.00)(150.00,120.00)(130.00,90.00)
\put(132,112){\vector(-1,-1){2}} 
\bezier{157}(90.00,80.00)(96.00,100.00)(106.00,95.00)
\put(90,80){\vector(-1,-1){2}} 
\bezier{132}(100.00,56.00)(130.00,40.00)(112.00,65.00)
\put(114,63){\vector(-1,1){2}} 
\bezier{137}(115.00,70.00)(125.00,60.00)(128.00,90.00)
\put(127,83){\vector(1,4){2}} 
\put(145.00,80.00){\makebox(0,0)[cc]{{\scriptsize Resurrected}}} 
\put(145.00,70.00){\makebox(0,0)[cc]{{\scriptsize $\gamma_{i}$}}} 
\put(143.00,60.00){\makebox(0,0)[cc]{{\scriptsize Prob $1-p_{i}-q_{i}-w_{i}$}}} 
\put(105.00,49.00){\makebox(0,0)[cc]{{\scriptsize $1-\gamma_{i}$}}} 
\put(47.00,70.00){\makebox(0,0)[cc]{{\scriptsize Dies}}} 
\put(47.00,60.00){\makebox(0,0)[cc]{{\scriptsize $\delta_{i}$}}} 
\put(50.00,99.00){\makebox(0,0)[cc]{{\scriptsize Infec}}}
\put(50.00,93.00){\makebox(0,0)[cc]{{\scriptsize Info}}}
\put(25.00,92.00){\makebox(0,0)[cc]{{\scriptsize Prob $p_{i}$}}}
\put(30.00,140.00){\makebox(0,0)[cc]{{\scriptsize $1-\delta_{i}$}}} 
\put(100.00,92.00){\makebox(0,0)[cc]{{\scriptsize Dies}}} 
\put(100.00,87.00){\makebox(0,0)[cc]{{\scriptsize $\delta_{i}$}}} 
\put(83.00,155.00){\makebox(0,0)[cc]{{\scriptsize Receives Virus}}}
\put(83.00,145.00){\makebox(0,0)[cc]{{\scriptsize $(1-\zeta_{i}(t))\nu_{i}$}}}
\put(120.00,99.00){\makebox(0,0)[cc]{{\scriptsize No}}}
\put(120.00,93.00){\makebox(0,0)[cc]{{\scriptsize Info}}}
\put(85.00,123.00){\makebox(0,0)[cc]{{\scriptsize Warn}}}
\put(85.00,119.00){\makebox(0,0)[cc]{{\scriptsize Info}}}

\put(135.00,122.00){\makebox(0,0)[cc]{{\scriptsize $\zeta_{i}(t)-\delta_{i}$}}}
\put(145.00,92.00){\makebox(0,0)[cc]{{\scriptsize Prob $q_{i}$}}}
\put(100.00,70.00){\makebox(0,0)[cc]{{\scriptsize Dead}}}
\bezier{139}(96.00,120.00)(105.00,130.00)(118.00,115.00)
\put(100,120){\vector(-2,-1){5}} 
\put(105.00,130.00){\makebox(0,0)[cc]{{\scriptsize Warn}}}
\put(105.00,115.00){\makebox(0,0)[cc]{{\tiny $(1-\zeta_{i}(t))$}}}
\put(101.00,110.00){\makebox(0,0)[cc]{{\tiny $\cdot(1-\nu_{i})$}}}
\bezier{149}(86.00,112.00)(100.00,90.00)(105.00,100.00)
\put(105,100){\vector(1,1){2}} 
\put(100.00,102.00){\makebox(0,0)[cc]{{\scriptsize $\chi_{i}$}}}
\bezier{159}(84.00,111.00)(78.00,90.00)(86.00,75.00)
\put(86,75){\vector(1,-1){2}} 
\put(63,95){\vector(-1,-1){2}} 
\bezier{179}(82.00,132.00)(85.00,143.00)(87.00,132.00)
\put(87,132){\vector(0,-1){2}} 
\put(95.00,135.00){\makebox(0,0)[cc]{{\tiny $1-\chi_{i}-\delta_{i}$}}}
\put(88.00,92.00){\makebox(0,0)[cc]{{\tiny $\delta_{i}$}}}
\put(69.00,118.00){\makebox(0,0)[cc]{{\tiny Prob $w_{i}$}}}
\put(100,20){\makebox(0,0)[cc]{Diagram 2: Transitions on each node}}
\end{picture}

Making the same node independence probability assumption that is stated in equation (\ref{indep}) and
taking into account the new states and transition probabilities shown in the Diagram 2, the equations (\ref{probp}) and (\ref{probq}) as well as the new equation corresponding to $w_{i}$ 
can be expressed as follows:
\begin{eqnarray}
p_{i}(t)&=&p_{i}(t-1)(1-\delta_{i})+q_{i}(t-1)(1-\zeta_{i}(t))\nu_{i} \label{probnp}\\
 &  & \nonumber\\
q_{i}(t)&=&q_{i}(t-1)(\zeta_{i}(t)-\delta_{i})+ \label{probnq}\\
	&  &(1-p_{i}(t-1)-q_{i}(t-1)-w_{i}(t-1))\gamma_{i} \nonumber\\
	&  &+\chi_{i}w_{i}(t-1) \nonumber \\
 &  & \nonumber\\
w_{i}(t)&=&(1-\zeta_{i}(t))(1-\nu_{i})q_{i}(t-1) \label{probw}  \\
    &  & +(1-\chi_{i}-\delta_{i})w_{i}(t-1) \nonumber 
\end{eqnarray}
Also in our case of study, instead of solving the Markov Chain, that is rather complicated, we will describe the behaviour of our system considering it as a dynamical system described by the equations (\ref{probnp}), (\ref{probnq}) and (\ref{probw}). Following  \cite{Hirsch} we will calculate the fixed points of the system. As we have stated before, in these  very points   the state probabilities become stable, then $p_{i}(t)=p_{i}(t-1)$, $q_{i}(t)=q_{i}(t-1)$ and $w_{i}(t)=w_{i}(t-1)$. Using (\ref{probnp}), (\ref{probnq}) and (\ref{probw}), we can state the following result

\begin{lemma}
$p_{i}(t)+q_{i}(t)+w_{i}(t) \rightarrow \frac{\gamma_{i}}{\gamma_{i}+\delta_{i}}$
\\
{\bf Proof:}
In the same way that has been done in \cite {Deepa1} we can do the subtraction $1-p_{i}(t)-q_{i}(t)-w_{i}(t)$ and simplify by renaming $x_{i}(t)=p_{i}(t)+q_{i}(t)+w_{i}(t)$ what give us the following linear system

\begin{equation}
x_{i}(t)=(1-\delta_{i}-\gamma_{i}) \cdot x_{i}(t-1)+\gamma_{i}
\end{equation}

In the fixed point $x_{i}(t)=x_{i}(t-1)$ so if we apply this to the last equation we  have that
\begin{equation}
x_{i}(t)=\frac{\gamma_{i}}{\gamma_{i}+\delta_{i}}
\end{equation}
that is, $p_{i}(t)+q_{i}(t)+w_{i}(t) = \frac{\gamma_{i}}{\gamma_{i}+\delta_{i}}$.
Then by Lemma 2 in \cite{Deepa1}, this convergence is exponential.

\end{lemma}

It is worthy to notice that the same results for the fixed points are obtained considering the linear behaviour of the system on these points. Using (\ref{probnp}), (\ref{probnq}) and (\ref{probw}) and for simplicity  dropping the indexes and the the time dependance, we  obtain the equations

\begin{equation}
\begin{array}{l}
-\delta \cdot p  +  \nu (1-\zeta) \cdot  q = 0\\
(-1 + \zeta  - \delta  - \gamma ) q + (-\gamma  + \chi ) w = -\gamma \\
(1 - \zeta  - \nu  + \zeta \nu) q + (1 - \chi  - \delta) w = 0\\
\end{array}
\end{equation}

Whose solutions can be obtained using Cramer's method 

\begin{equation}
\begin{array}{l} \label{fixpoint3}
p =-\frac{\gamma \cdot (-1+\zeta)\nu (\delta +\chi )}{(\gamma+\delta) \left(\delta +\delta ^2-\delta  \zeta +\delta  \chi +\nu  \chi
-\zeta  \nu  \chi \right)} \\
\\
q =\frac{\gamma  \delta  (\delta +\chi )}{(\gamma +\delta ) \left(\delta +\delta ^2-\delta  \zeta +\delta  \chi +\nu  \chi
-\zeta  \nu  \chi \right)}\\
\\
w=\frac{\gamma  \delta  (1-\zeta -\nu +\zeta  \nu )}{(\gamma +\delta ) \left(\delta +\delta ^2-\delta  \zeta +\delta  \chi
+\nu  \chi -\zeta  \nu  \chi \right)}

\end{array}
\end{equation}
 
 Once more, this expression can be simplified if we observe that the stable state \emph{No Info} is related with the desired \emph{fast extinction} condition that is also related with Markov chain probability condition $(1-\zeta)\rightarrow 0$, that implies $p \rightarrow 0$, this can be summarized as 
 
\begin{equation}
\begin{array}{l} \label{fixpoint4}
p =0 \\
\\
q =\frac{\gamma }{\gamma +\delta }\\
\\
w=0

\end{array}
\end{equation} 
 

Given that by hypothesis, the probability of events in each node are independent, we can try to analyse the problem using the Markov approach. This will be made in the next subsection.

\subsection{Markovian Analysis}
In this section we analyse again the ergodicity behaviour of the Markov chains associated inside each node under our model. 

The calculation steps performed were :
\begin{itemize}

\item We obtain the transition matrix $P$ 
{ \scriptsize
\begin{equation}
P=
\left (
\begin{array}{ccccc} 

\zeta_{i}(t)-\delta_{i} & (1-\zeta_{i}(t))\nu_{i} & (1-\zeta_{i}(t))(1-\nu_{i})          & \delta_{i} \\ 
        0               & 1-\delta_{i}            &         0                            & \delta_{i} \\ 
        \chi_{i}        &                   0     & 1-\chi_{i}-\delta_{i}           & \delta_{i} \\  
        \gamma_{i}      &   0                    &       0                      & 1-\gamma_{i} 

\end{array}
\right )
\end{equation}
}
\\
\item we calculate the Z-transform of the matrix $M=I-zP$, that is
{ \scriptsize
\begin{equation}
M=
\left(
\begin{array}{cccc}
 1-z (-\delta +\zeta ) & z (-1+\zeta ) \nu  & z (-1+\zeta ) (1-\nu ) & -z \delta  \\
 0 & 1-z (1-\delta ) & 0 & -z \delta  \\
 -z \chi  & 0 & 1-z (1-\delta -\chi ) & -z \delta  \\
 -z \gamma  & 0 & 0 & 1-z (1-\gamma )
\end{array}
\right)
\end{equation}
}
\item The inverse matrix $M^{-1}$ of $M$ is obtained

\item The inverse Z-transform is applied to $M^{-1}$, obtaining 

$H(n)={\cal{Z}}^{-1}\{{M^{-1}}\}$

This equation can be written as a sum of  two matrices \cite{Howard}, $S$ that corresponds to the steady behavior  and $T$ that represents the transient behavior of the Markov Chain, that is  

$H(n)=k_1S+k_2(C_1)^{n}T$,

where $k_1$ and $k_2$ are constants. Based on this procedure we obtained the ergodicity condition, that is, $C_1 <1$ what assures the convergence to a steady state no matter what the initial state was. 

\noindent

All the calculations were made using MATHEMATICA and the ergodicity condition expression obtained is given by
{\footnotesize
\begin{equation}
C_1=\frac{1-2 \delta +\zeta -\chi +\sqrt{1-2 \zeta +\zeta ^2+2 \chi -2 \zeta  \chi -4 \nu  \chi +4 \zeta  \nu  \chi +\chi ^2}}{\delta^2+\zeta +(-1+\nu ) \chi -\zeta  \nu  \chi +\delta  (-1-\zeta +\chi )}.
\end{equation}
}
\end{itemize}

Again, it can be noticed from the calculation above, that the ergodicity of the associated Markov chain depend on the choice of the transition probabilities involved. It should be also mentioned that the algebraic manipulations become even more complex than in the case of the Markov chain of \cite{Deepa1} given that our Markov chain have one state more. 

\subsection{Jacobian and fix point}

In our case of study, we can proceed as \cite{Deepa1}. Firstly, let us define the column vectors $\vec{\mathbf{p}}(t)=(p_1(t),p_2(t), \dots, p_N(t))$, $\vec{\mathbf{q}}(t)=(q_1(t),q_2(t), \dots, q_N(t))$ and $\vec{\mathbf{w}}(t)=(w_1(t),w_2(t), \dots, w_N(t))$. Let the  vector $\vec{\mathbf{v}}(t)=(\vec{\mathbf{p}}(t),\vec{\mathbf{q}}(t), \vec{\mathbf{w}}(t))$ be the concatenation of the previous vectors and let $\vec{\mathbf{v}}_f(t)$ be the vector $\vec{\mathbf{v}}(t)$ evaluated at the fixed point. Then, the entire system can be described  by
\begin{equation}
\vec{\mathbf{v}}(t)=\mathbf{f}(\vec{\mathbf{v}}(t-1))
\end{equation}
where 
{\footnotesize
\begin{displaymath}
f_i(\vec{\mathbf{v}}(t-1))= \left \{
\begin{array}{lllll}
	p_{i}(t-1)(1-\delta_{i}) &~~\text{if~} i\leq N &\\
	+q_{i}(t-1)(1-\zeta_{i}(t))\nu_{i} \\
  &\\	
	q_{i}(t-1)(\zeta_{i}(t)-\delta_{i}) \\
	+(1-p_{i}(t-1)-q_{i}(t-1)   &~~\text{if~} N<i\leq 2N &\\
	-w_{i}(t-1))\gamma_{i}+\chi_{i}w_{i}(t-1) \\
  &\\
  (1-\zeta_{i}(t))(1-\nu_{i})q_{i}(t-1)\\
  +(1-\chi_{i}-\delta_{i})w_{i}(t-1) &~~\text{if~} 2N<i\leq 3N &
\end{array}
\right .
\end{displaymath}
}

Now, following \cite{Hirsch},  let us define the  the Jacobian matrix of the system,  $\nabla \mathbf{f}$  as
\begin{displaymath}
[  \nabla \mathbf{f} ]_{ij}=\frac{\partial f_i(\vec{\mathbf{v}}(t-1))}{\partial \vec{\mathbf{v}}(t-1) }.
\end{displaymath}

In  order to explore the  asymptotic stability of fixed points and according to \cite{Hirsch} we will have to take into account the value of the function $\nabla \mathbf{f}$  in these points  In our case,  we obtain the  $3N \times 3N$ Jacobian matrix

\begin{equation}
\nabla(\mathbf{f})|_{\vec{\mathbf{v}}_{f}}= \left [
\begin{array}{lllll}
S      & |  & 0 & |  & 0 \\
\hline 
S_{1}  & | & S_{2} & | & S_{3} \\
\hline 
S_{4}  & | & 0 &| & S_{6}
\end{array}
\right ]
\end{equation} 

where each block  $S, S_{1},\dots $ is a  $N \times N$ matrix whose elements are given by 

\begin{equation}
S_{ij}= \left \{
\begin{array}{ll}
	1-\delta_{i}  &~~~~\text{if~} i=j \\
	r_{j} \beta_{ji} \nu_{i}\frac{\gamma_{i}}{\gamma_{i}+\delta_{i}} & ~~~~\mbox{otherwise}
\end{array}
\right .
\end{equation}

and the others are

\begin{equation}
S_{1ij}= \left \{
\begin{array}{ll}
	-\gamma_{i}  &~~~~\text{if~} i=j \\
	-r_{j} \beta_{ji} \frac{\gamma_{i}}{\gamma_{i}+\delta_{i}} & ~~~~\mbox{otherwise}
\end{array}
\right .
\end{equation}

\begin{equation}
S_{2ij}= \left \{
\begin{array}{ll}
	1-\gamma_{i}-\delta_{i}  &~~~~\text{if~} i=j \\
	0 & ~~~~\mbox{otherwise}
\end{array}
\right .
\end{equation}

\begin{equation}
S_{3ij}= \left \{
\begin{array}{ll}
	-\gamma_{i}+\chi_{i}  &~~~~\text{if~} i=j \\
	0 & ~~~~\mbox{otherwise}
\end{array}
\right .
\end{equation}

\begin{equation}
S_{4ij}= \left \{
\begin{array}{ll}
	0  &~~~~\text{if~} i=j \\
	r_{j} \beta_{ji} \frac{(1-\nu_{i})\gamma_{i}}{\gamma_{i}+\delta_{i}} & ~~~~\mbox{otherwise}
\end{array}
\right .
\end{equation}

\begin{equation}
S_{6ij}= \left \{
\begin{array}{ll}
	1-\chi_{i}-\delta_{i}  &~~~~\text{if~} i=j \\
	0 & ~~~~\mbox{otherwise.}
\end{array}
\right .
\end{equation}

Once we have calculated the Jacobian matrix of the system we can extend the  results of {\bf Lemma 2}. That is, if the  largest eigenvalue (in magnitude) is less than one then it is assured that the system is asymptotically stable in the fixed  
point $\vec{\mathbf{v}}$ and the dynamical system will exponentially  tend to the fixed point whatever was the initial state.
Those interested in the detailed proof of this {\bf Lemma 2} can consult the appendix of \cite{Deepa1}.

\section{Simulations.}
 Until now, we have theoretically described the behaviour of the net.  
\begin{figure}[hbtp]
\caption{Number of carriers $C(t)$ vs time (simulation epochs) in Chakrabarti model under the threshold. Values for $\delta=0.1$ and $\gamma=0.01$}
\label{ChakraUnder}
\includegraphics[width=\linewidth]{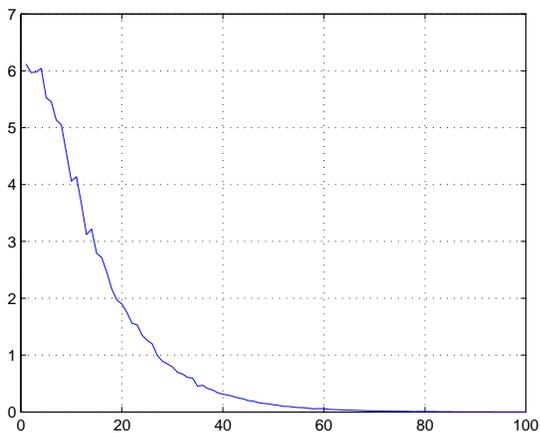}
\end{figure}

\begin{figure}[hbtp]
\caption{Number of carriers $C(t)$ vs time (simulation epochs) in our model under the threshold. Values for $\delta=0.1$ and $\gamma=0.01$.}
\label{OurUnder}
\includegraphics[width=\linewidth]{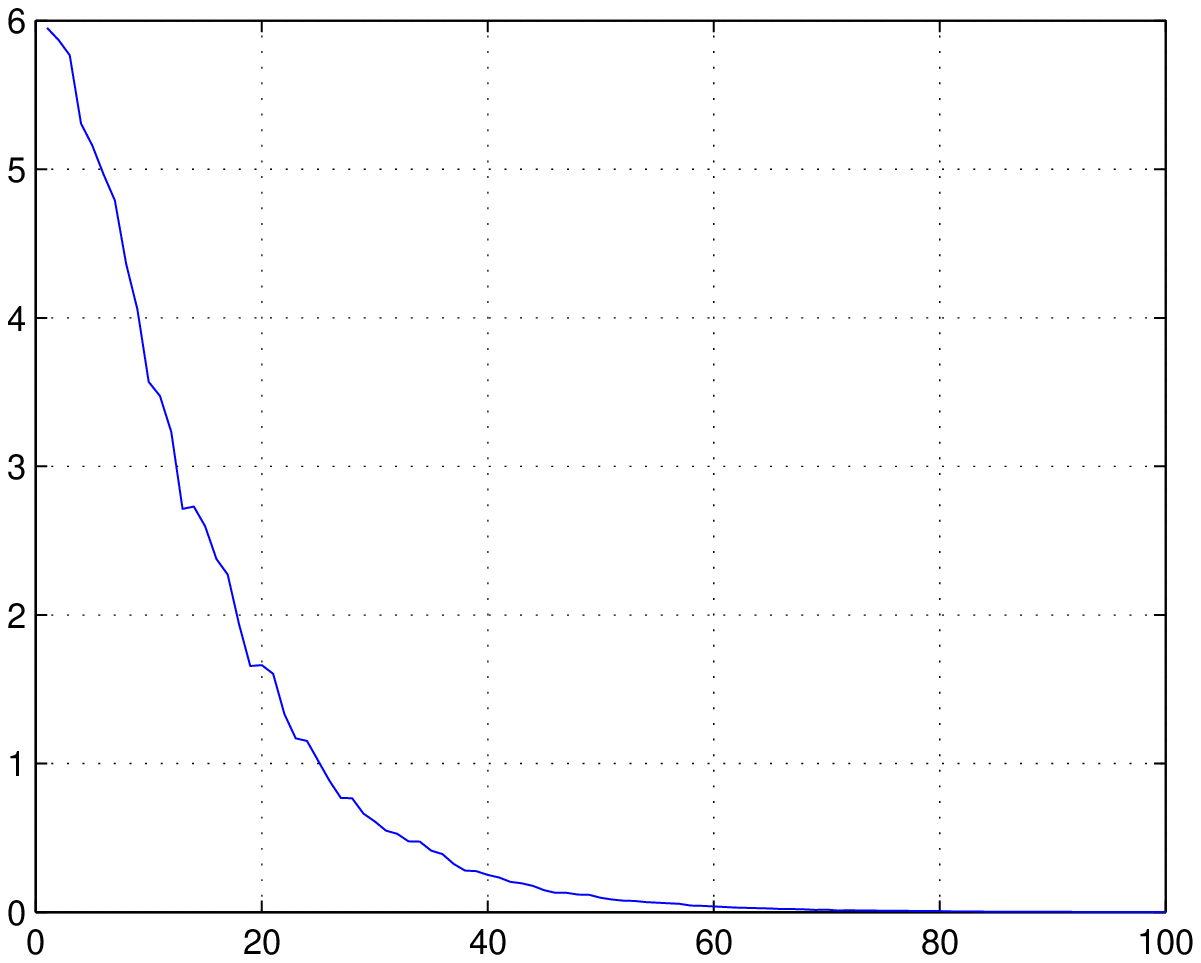}
\end{figure}
\begin{figure}[hbtp]
\caption{Number of carriers $C(t)$ vs time (simulation epochs) in Chakrabarti model on the threshold. Values for $\delta=0.07$ and $\gamma=0.004$}
\label{ChakraOn}
\includegraphics[width=\linewidth]{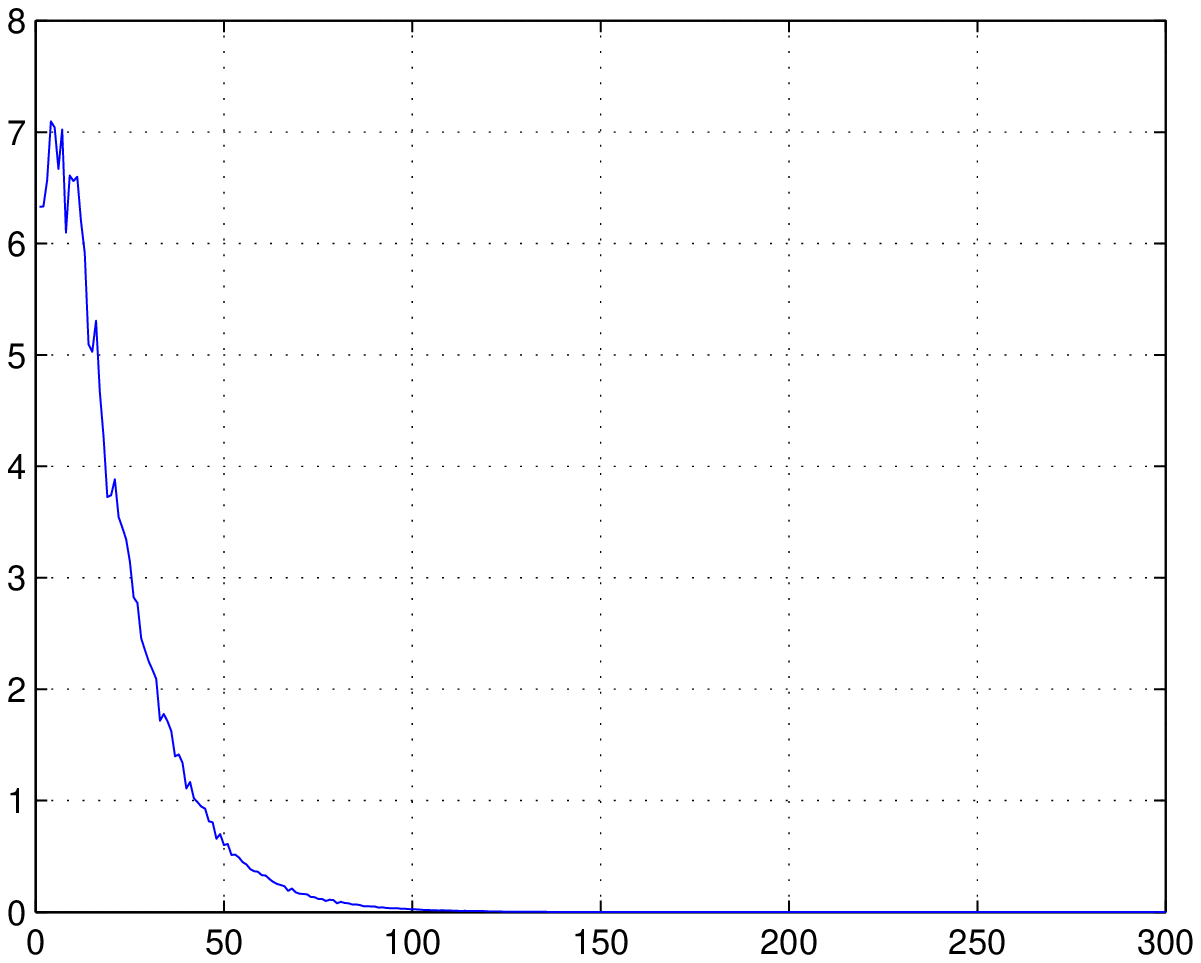}
\end{figure}

\begin{figure}[hbtp]
\caption{Number of carriers $C(t)$ vs time (simulation epochs) in our model on the threshold. Values for  $\delta=0.07$ and $\gamma=0.004$.}
\label{OurOn}
\includegraphics[width=\linewidth]{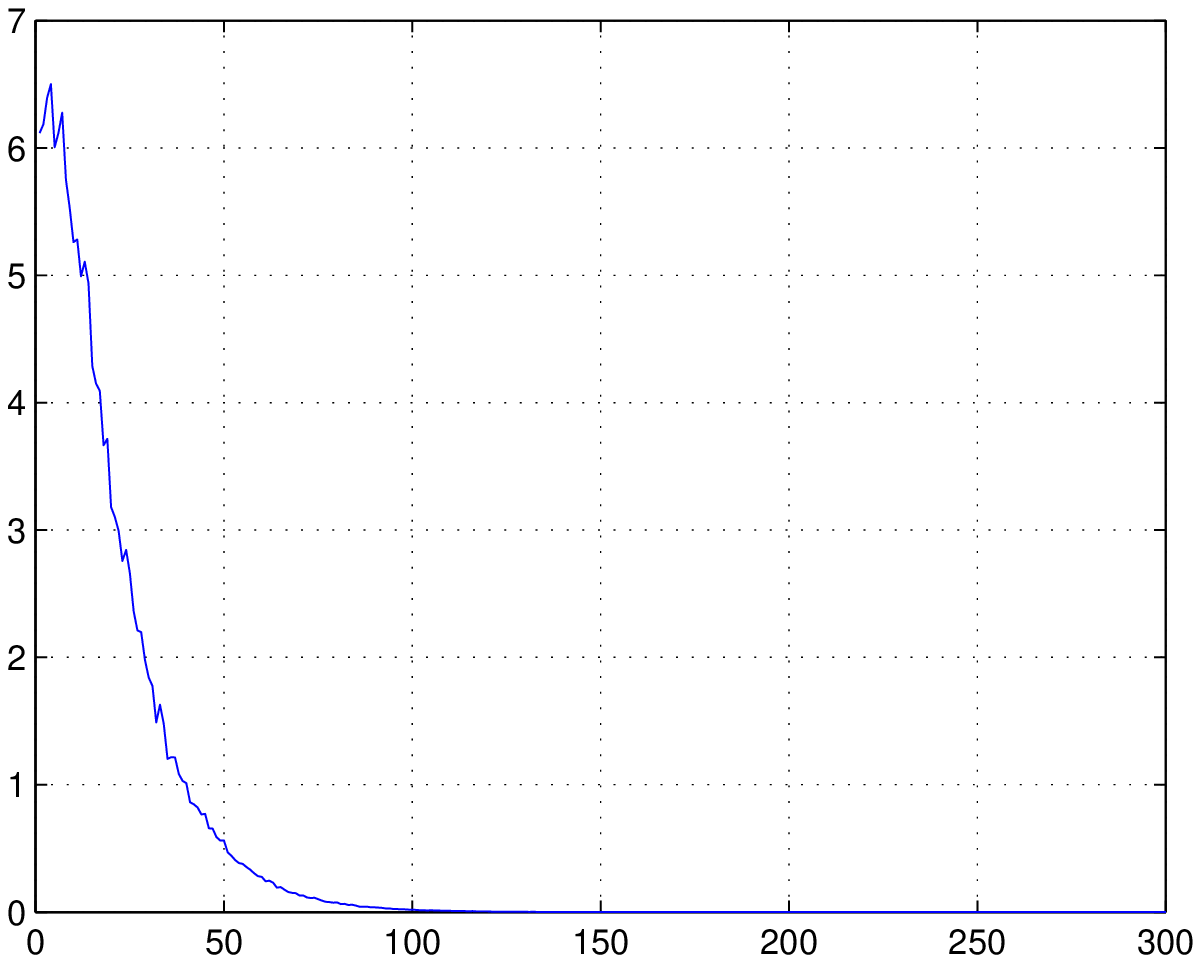}
\end{figure}

\begin{figure}[hbtp]
\caption{Number of carriers $C(t)$ vs time (simulation epochs) in Chakrabarti model above the threshold. Values for $\delta=0.01$ and $\gamma=0.01$}
\label{ChakraAbove}
\includegraphics[width=\linewidth]{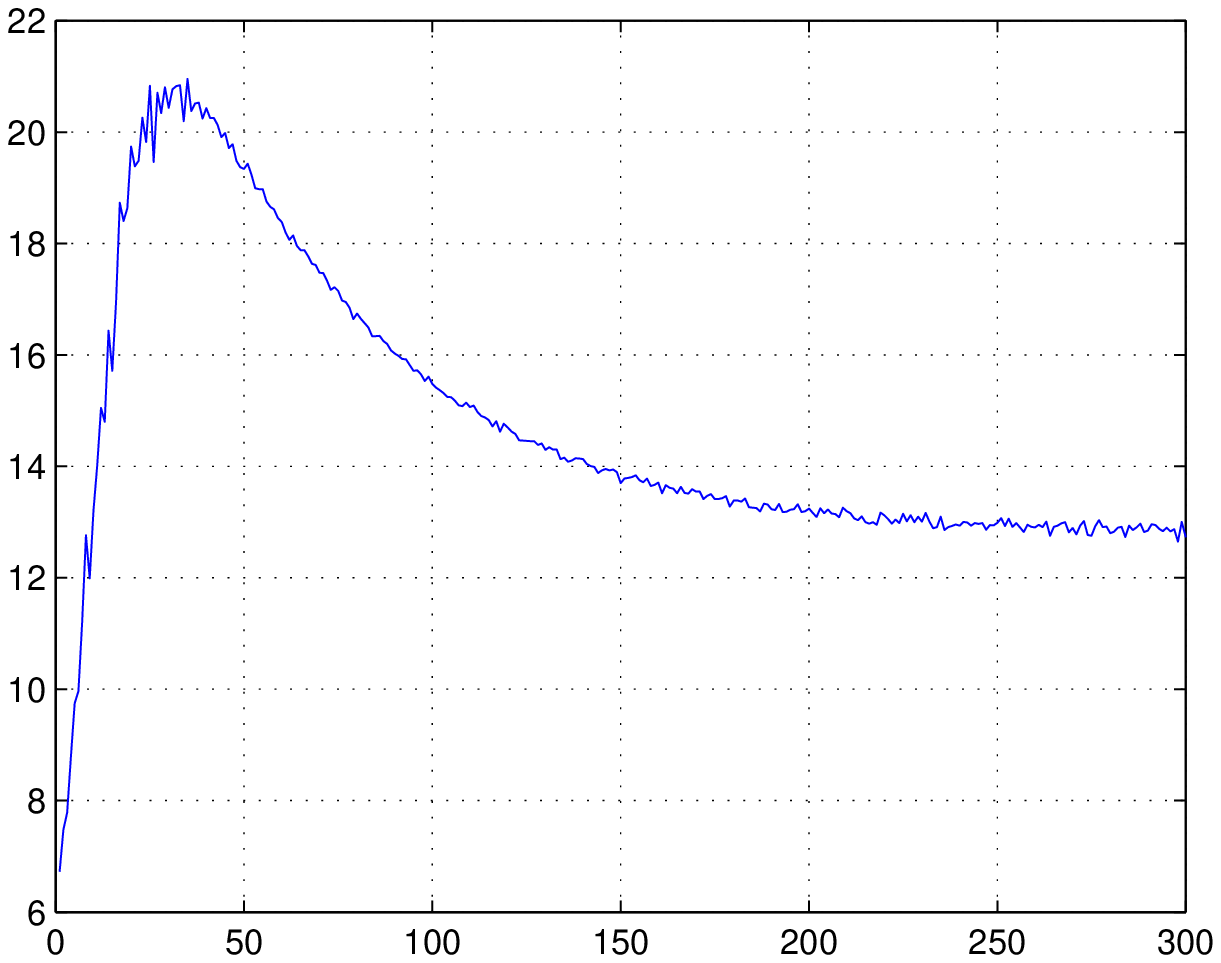}
\end{figure}

\begin{figure}[h!]
\caption{Number of carriers $C(t)$ vs time (simulation epochs) in our model above the threshold. Values for $\delta=0.01$ and $\gamma=0.01$}
\label{OurAbove}
\includegraphics[width=\linewidth]{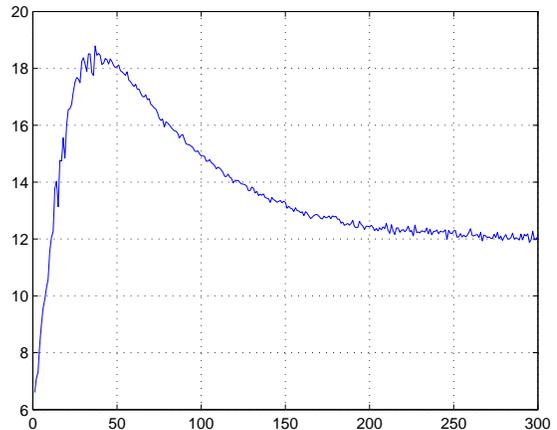}
\end{figure}
\setcounter{topnumber}{2}
Using the Dynamical Systems approach we have been able to predict the conditions (a threshold) under which fast extinction is reached by using a limited set of parameters that assures the we will converge to a fixed point.

In order to complete the study of our model and compare its performance with the Chakrabarti model, we have made different simulations corresponding to the cases in which we are under, on and above  the threshold values established in the previous section.

 For this end, we have randomly generated the adjacency matrix  corresponding to a thirty node graph. 
We have taken, for the sake of comparison,  the same set of parameters that appears in section 4 of \cite{Deepa1}, that is, $r=0.1$ and $\beta_{i,j}=0.1 $. 
Additionally, we used the different values of $\delta$ and $\gamma$ proposed in \cite{Deepa1}  corresponding to P2P GNUTELLA data sets. We have fixed $\nu=0.8$ and $\chi=0.1$ and  we have started with six infected  nodes that we choose randomly. The number of time steps have been fixed in our simulation to one and three hundred.

In Figures \ref{ChakraUnder} and \ref{OurUnder} it is shown that if the settings of the parameters values  fulfil the  fast extinction condition  then the same result is obtained in both models.  We can have a diverse set  of parameters values as long as we are under the threshold condition for achieving  fast extinction. 

In Figures \ref{ChakraOn} and \ref{OurOn} we show that if the set of parameters are combined in such a way that they  give exactly the  threshold value the fast extinction is achieved again in both models but in this case the fast extinction is slower than in the first case.
When the parameters values are combined in such a way that we are beyond the  threshold value then fast extinction is no longer accomplished and the number of carriers grow very fast and eventually the whole set of nodes could become infected. This behaviour is shown in Figures \ref{ChakraAbove} and \ref{OurAbove}. In this case it can be observed a slight difference between both models  behaviour due to the presence of the parameter $\nu_i$.
\FloatBarrier

\section{Conclusions and Future Work}
As we have exposed in the sections corresponding to our proposal, if under our model $1-\zeta_{i}(t) \rightarrow 0$, our \emph{fix point and fast extinction condition} are consistent with those obtained in \cite{Deepa1}. In the other side, if under our model $1-\zeta_{i}(t) > 0$, then the \emph{fix point and fast extinction condition} mentioned in the appendix section of \cite{Deepa1} are not longer valid. In this last case the dynamical system becomes nonlinear, and the degree of non-linearity will depend on the topology of the network. In this new setting our $\nu_{i}$ parameter start to play a r\^{o}le in the virus as well as antidote spreading on the network. In the future  we will study this problem.      
If we take as starting point this scenario it can be interesting to ask if the system falls in a chaotic regime and if this is the case then how the stability of the network can be  re-established. If we want to achieve this state of the system we can recall the synchronization and chaos tools developed in the research field of automatic control. This is one of the subjects that we will try to explore in the future.



\begin{thebibliography}{10}

\bibitem{Barab0}
R.~Albert and A.L. Barab\'asi.
\newblock Error and attack tolerance of complex networks.
\newblock {\em Nature}, 406, 2000.

\bibitem{Barab1}
A.L. Barab\'asi and R.~Albert.
\newblock Emergence of scaling in random graphs.
\newblock {\em Science}, 286:509--512, 1999.

\bibitem{chayes1}
Ayalvadi~Ganesh Christian~Borgs, Jennifer~Chayes and Amin Saberi.
\newblock How to distribute antidote to control epidemics.
\newblock {\em Random Structures and Algorithms John Wiley and Sons, Inc. New
  York, NY, USA}, (Volume 37, Issue 2):204--222, 2010.

\bibitem{Durrett1}
R.~Durrett and X.-F. Liu.
\newblock The contact process on a finite set.
\newblock {\em The Annals of Probability}, 16(3):1158--1173, 1988.

\bibitem{Radicchi1}
Alain~Barrat Filippo~Radicchi, Jose J.~Ramasco and Santo Fortunato.
\newblock Complex networks renormalization: Flows and fixed points.
\newblock {\em Physical Review Letters}, (Volume 65):1487011--1487014, 2008.

\bibitem{Hirsch}
M.~W. Hirsch and S.~Smale.
\newblock {\em Differential Equations,Dynamical Systems, and Linear Algebra}.
\newblock Academic Press, second edition, 1974.

\bibitem{Howard}
Ronald~A. Howard.
\newblock {\em Dynamic Programming and Markov Processes}.
\newblock The MIT Press, fourth edition, 1966.

\bibitem{Deepa1}
C.~Faloutsos S. Madden C.~Guestrin J.~Leskovec, D.~Chakrabarti and M.~Faloutsos.
\newblock Information survival threshold in sensor and p2p networks.
\newblock In {\em IEEE INFOCOM 2007}, 2007.

\bibitem{Kempe1}
D.~Kempe and J.~Kleinberg.
\newblock Protocols and impossibility results for gossip-based communication
  mechanisms.
\newblock In {\em Proceedings of the Symposium on Foundations of Computer
  Science (FOCS 2002)}, 2002.

\bibitem{Falou1}
P.~Faloutsos M.~Faloutsos and C.~Faloutsos.
\newblock On power-law relationships of the internet topology.
\newblock In {\em In Proceedings Sigcomm 1999}, 1999.

\bibitem{Barab2}
D.~Ben-Avraham A.L.~Barab\'asi N.~Schwartz, R.~Cohen and S.~Havlin.
\newblock Percolation in directed scale-free networks.
\newblock {\em Physical Review E}, 66(1):0151041--0151044, 2002.

\bibitem{Alon1}
Itai~Benjamini Noga~Alon and Alan Stacey.
\newblock Percolation on finite graphs and isoperimetric inequalities.
\newblock {\em The Annals of Probability}, (Volume 32, No.3A):1727--1745, 2004.

\bibitem{Satorras1}
Romualdo Pastor-Satorras and Alessandro Vespignani.
\newblock Epidemic dynamics and endemic states in complex networks.
\newblock {\em Physical Review E}, (Volume 63, Issue 2):0661171--0661178, 2001.

\bibitem{Satorras2}
Romualdo Pastor-Satorras and Alessandro Vespignani.
\newblock Epidemic spreading in scale-free networks.
\newblock {\em Physical Review Letters}, (Volume 86, Number 14):3200--3203,
  2001.

\bibitem{Satorras3}
Romualdo Pastor-Satorras and Alessandro Vespignani.
\newblock Epidemic dynamics in finite size scale-free networks.
\newblock {\em Physical Review E}, (Volume 65):0351081--0351084, 2002.

\bibitem{Bollobas1}
B\'ela~Bollob\'asi Paul~Balister.
\newblock Bond percolation with attenuation in high dimensional voronoi
  tilings.
\newblock {\em Random Structures and Algorithms, Wiley InterScience}, pages
  5--10, 2009.

\end{thebibliography}

\end{document}